# Light repolarization by scattering media

J. Sorrentini, M. Zerrad, G. Soriano and C. Amra

*Institut Fresnel, UMR CNRS 6133, Universités d'Aix-Marseille, Ecole Centrale Marseille, Faculté des Sciences et Techniques de Saint-Jérôme, 13397 Marseille Cedex 20, France*

*claude.amra@fresnel.fr*

The polarization of a coherent depolarized incident light beam passing through a disordered medium is investigated. The local polarization of the scattered far field and the probability density function are calculated and show an excellent agreement with experiment. It is demonstrated that complex media may confer high degree of polarization (0.75 DOP average) to the incident unpolarized light.

The state of polarization is one of the main observable parameters of an optical field. Many practical situations exist that make the light polarization properties depend on the spatial location. Indeed the state polarization of a light beam [1-3] will change by propagation in free-space [4, 5], by propagation in turbulent atmosphere [6, 7], by beam combination [8], after scattering by a rough surface [9-14] or an inhomogeneous medium [15-19].

Most of these works are devoted to the loss of polarization that can take place on the incident light, considering a full polarization but different spatial and temporal coherence properties for the incident beam. Different formalisms were proposed including Mueller-Stokes [17], cross spectral density matrices [7] and electromagnetic theories. Such loss of polarization (or depolarization process) most often originates from a temporal average of uncorrelated polarization modes of the optical field [6, 7, 11, 15, 17, 18, 20], though spatial average may also be responsible for depolarization of a fully polarized incident beam [9, 10, 12, 13, 16, 21] when the state of polarization rapidly varies within the detection area.

Scattering by arbitrary inhomogeneous media is known to modify the polarization or depolarization properties of the illumination beam. Usually the incident polarization of a light beam is lost after scattering by a highly inhomogeneous medium, which reduces the interest of polarimetric techniques to probe random media [13]. However one can have the benefits of a reversible effect in the sense that the same media may allow to significantly increase the polarization degree of a fully depolarized incident light. This is the scope of this paper where it is shown that unpolarized light can be "ordered" by a complex scattering process.

Repolarization of light has been observed by different authors; in particular Mujat and Dogariu [8] used beam combination inside an interferometer and emphasized a procedure to produce partial polarization at the system output, though the input was unpolarized light. In this work similar results are obtained with light scattering in the far field, though the scattering process is strongly different from that of specular beams.

A phenomenological approach is first used to calculate the spatial repartition of the local Degree of Polarization (DOP) of unpolarized light after transmission by a random medium and propagation in air. The average value and the probability density function (pdf) of the DOP are investigated and an excellent agreement is obtained between numerical and experimental results. The high average polarization degree of light ($\approx 75\%$) compared with the incident one (<4%) allows considering that light has been *ordered* when passing through the disordered medium.

Let us consider a fully coherent and fully depolarized incident light beam characterized by the electric field $E(r,t)$ illuminating a scattering medium whose Jones matrix is denoted $M = (v_{uv})$, and $r$ is the spatial coordinate. In the plane $z = z_0$ (see Fig. 1), the field is written as:

$$E(r,t) = \sqrt{I(r)} \begin{pmatrix} e_s(t) \\ e_p(t) \end{pmatrix} \qquad (1)$$

Where $\sqrt{I(r)}$ and $\begin{pmatrix} e_s(t) \\ e_p(t) \end{pmatrix}$ describe spatial and temporal variations. The Degree of Polarization of $E(r,t)$ is assumed to be zero whatever the r location. Therefore, at any point of the plane $z=z_0$, no temporal correlation exists between the Transverse Electric (*TE* or *s*) and the Transverse Magnetic (*TM* or *p*) modes [17], that is :

$$\mu = \frac{\left\langle e_s(t)\overline{e_p(t)}\right\rangle}{\sqrt{\left\langle |e_s(t)|^2\right\rangle \left\langle |e_P(t)|^2\right\rangle}} = \left\langle e_s(t)\overline{e_p(t)}\right\rangle = 0 \qquad (2)$$

with bars denoting the complex conjugaison. In this relation $e_S(t)$ and $e_P(t)$ are normalized as: $<e_S(t)> = <e_P(t)> = 1$. The brackets <> stand for the temporal average. The spectral bandwidth $\Delta\omega$ of $E(r,t)$ is centered on the average frequency $\omega_0$ and matches the quasi-monochromatic condition: $\Delta\omega/\omega_0 \ll 1$. Moreover this beam illuminates a scattering medium whose linear response is not frequency-dependent within the spectral domain, in order to preserve temporal coherence.

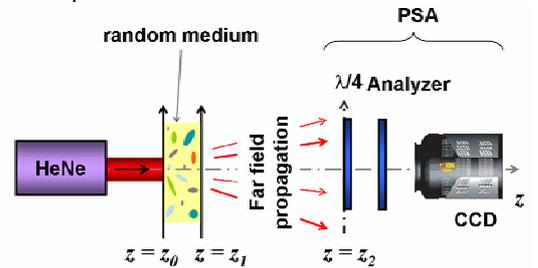

Fig. 1 : Schematic view of the experiment

Therefore, following the schematic view of Fig. 1, one can write the field $E^{sc}$ scattered in the far field at infinity and at direction $(\theta,\phi)$ as:

$$E^{sc} = \begin{pmatrix} v_{ss}e_s(t) + v_{ps}e_p(t) \\ v_{sp}e_s(t) + v_{pp}e_p(t) \end{pmatrix} \quad (3)$$

Where the scattering coefficients $(v_{uv})$ take into account the angular variations of the v-polarized scattered wave resulting from a u-polarized illumination. Notice here that each angular direction corresponds in the far field to a particular location in a plane perpendicular to propagation. The $(v_{uv})$ coefficients can be predicted with exact electromagnetic methods [22-27] that give quantitative values [13]. Notice that these methods take into account the whole illuminated area on the sample under study; moreover, because the complex medium is perfectly identified the question of averaging the electromagnetic calculation over multiple realizations would be irrelevant. However these numerical techniques are highly time-consuming for 3D arbitrary bulk structures and may not converge. For this reason we used a fully developed speckle model [28] to predict the statistical behaviour of the $(v_{uv})$ matrix. Within this approach and considering a bulk scattering process, the four $v_{ij}$ terms are known [12] to be mutually uncorrelated for a lambertian sample and to have similar average speckle patterns.

The Degree of Polarization (DOP) is defined from the coherence matrix in [29, eq. 4.3-36, p136]. Let us now express the DOP of the scattered field $E^{sc}$ as a function of the correlation $\mu^{sc}$ between its polarization modes:

$$DOP^{sc} = \sqrt{1 - 4\beta\left(1 - |\mu^{sc}|^2\right)/(1+\beta)^2} \quad (4)$$

with β the polarization ratio:

$$\beta = \frac{\langle |v_{SS}e_S(t) + v_{PS}e_P(t)|^2 \rangle}{\langle |v_{SP}e_S(t) + v_{PP}e_P(t)|^2 \rangle} \quad (5)$$

and the correlation:

$$\mu^{sc} = \frac{\langle (v_{SS}e_S(t) + v_{PS}e_P(t))\overline{(v_{SP}e_S(t) + v_{PP}e_P(t))} \rangle}{\sqrt{\langle |v_{SS}e_S(t) + v_{PS}e_P(t)|^2 \rangle \langle |v_{SP}e_S(t) + v_{PP}e_P(t)|^2 \rangle}} \quad (6)$$

Provided that all media are static (the scattering coefficients are time constants), Eq. (2) allows to write:

$$\beta = \frac{|v_{SS}|^2 + |v_{PS}|^2}{|v_{SP}|^2 + |v_{PP}|^2} \quad (7)$$

and $\mu^{sc} = \dfrac{v_{ss}\overline{v}_{sp} + v_{ps}\overline{v}_{pp}}{\sqrt{(|v_{ss}|^2 + |v_{ps}|^2)(|v_{sp}|^2 + |v_{pp}|^2)}} \quad (8)$

Because the $(v_{uv})$ coefficients are independent in the general case of arbitrary random media, Eq. (8) ensures that $\mu^{sc}$ will not be identically equal to zero. So, even though the illumination beam is perfectly un-polarized, the scattered light may be partially or totally polarized depending on the space location.

Numerical simulations have been performed to illustrate this phenomenon. Each speckle pattern $(v_{uv})$ is obtained via the Fourier Transform of a random phasor matrix [28]. Here, the non-zero domain is a square of $2^7$ points length within a square of $2^{10}$ points length. Fig. 2 shows the spatial repartition of the local DOP of the scattered far field at infinity in a (x,y) plane perpendicular to propagation ($z=z_2$ in Fig. 1). The pdf (probability density function) DOP function follows a $p(u) = 3u^2$ law as shown in Fig. 4. The resulting average of local DOP is 0.75, as given by:

$$\int_0^1 up(u)du = 3/4 \quad (9)$$

Such value emphasizes a significant increase of polarisation. Notice that the pdf function and its average are here deduced from numerical simulation and not by theoretical analysis of the statistical properties of the scattering process.

Eq. (9) indicates that light scattered by a highly inhomogeneous bulk under unpolarized illumination will exhibit a 75% average polarization rate. In other terms, polarization modes have recovered partial order when passing through the disordered medium.

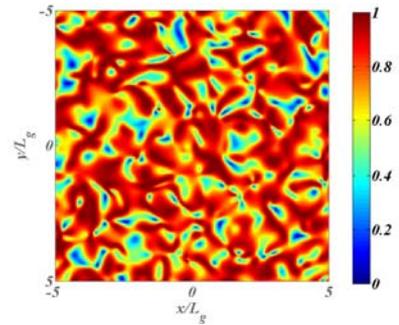

Fig. 2 : Calculation of the local DOP with a random phasor matrix. The resulting DOP average is 0.75. $L_g$ is the mean speckle size.

To go further, experiment has been used to confirm the "re-polarization" of unpolarized light by a scattering medium. A highly inhomogeneous bulk (a scattering calibration sample made of $MgF_2$, with 100% scattering and a lambertian pattern) is illuminated with a collimated He-Ne ($\lambda = 632.8$ nm) unpolarized (incident DOP ≈ 4%) laser beam of 3 mm diameter. The mean speckle size at the 1m distance associated to the measurement is Lg=0,2mm. The local DOP of the light scattered in the far field is classically measured [29] via the four Stokes images measurement. No lens is present in the system. The optical elements of the PSA are a quarter wave plate, a linear analyzer and a high sensitivity 1024*1024 pixels CCD array.

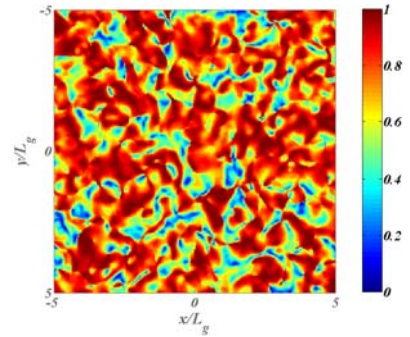

Fig. 3: Measurement of the local DOP. The resulting average is 0.75.

Fig. 3 shows the transverse variations of the DOP recorded in the plane $z=z_2$. Again the measured average of the DOP is 0.75, and the pdf law follows $3u^2$, in excellent agreement with prediction (see Fig. 4). Furthermore, this result is intrinsically related to the random phasor model [28], and thus should hold for most strongly disordered media.

At this step, one can say that calculation and measurements are in excellent agreement to emphasize the process of light repolarization by scattering media. An illustration was given with a highly inhomogeneous bulk and the result is a 0.75 average degree of polarization and a $3u^2$ pdf probability DOP function.

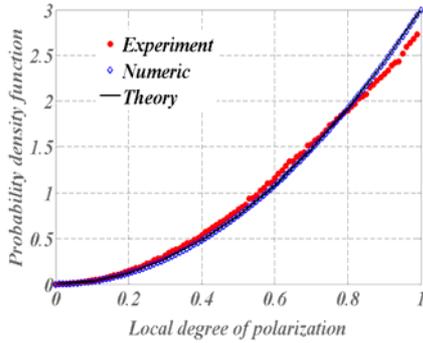

Fig. 4 : Calculation and measurement of the probability density function of the DOP. Both curves follow a $p(u)=3u^2$ variation.

In a more general way one may wonder whether specific media could allow to confer full polarization to unpolarized light. Following relation (8), one can show that such media would exhibit scattering coefficients following the condition:

$$\text{DOP} = 1 \Leftrightarrow \nu_{SS}\nu_{PP} = \nu_{SP}\nu_{PS} \qquad (10)$$

Provided that relation (10) is satisfied, light scattering would be fully polarized in whole space, despite the fact that the incident DOP is near zero. Solving this last equation addresses inverse problems that are outside the scope of this paper, but that justify additional efforts to search for the existence of solutions.

It is also necessary to notice one key difference in the repolarization processes obtained by beam combination inside an interferometer [8] and by light scattering. In the first situation and though the beams are combined, there is no mixing (S with P) of the polarization modes, that is, only the S modes (resp. P modes) are superimposed for each beam. Therefore the modes cross-correlation is not changed (remains equal to zero) so that temporal disorder is not reduced. The repolarization process only results from the relative weight of energy carried on each axis, which was modified by the interferometer; to be complete, in this interferometer experiment repolarization is connected to the polarization ratio β and vanishes in the case β = 1, due to the relationship:

$$\mu = 0 \Leftrightarrow DOP = |1-\beta|/|1+\beta| \qquad (11)$$

On the other hand, light scattering allows a spontaneous mixing of the polarization modes (see relation (3)), due to the presence of cross-scattering coefficients. Such mixing of S and P modes describes a linear combination of random variables (the polarization modes) on each axis. Hence the resulting variables may exhibit high cross-correlation, though the initial ones were totally uncorrelated. The temporal disorder is reduced and repolarization occurs. This result is valid whatever the β value.

We also notice that the repolarization process induced by scattering would vanish in the absence of cross-scattering coefficients. This is the reason why low-level scattering do not repolarize light in the incidence plane, since perturbative theories [22, 30] predict these coefficients to be zero.

At last, results similar to the scattering repolarization can be recovered by different techniques. Anisotropic materials would provide similar effects due to cross-polarized terms. Also, beam focusing allows a repolarization effect [31]; in this last situation, the linear combination of random variables appears within a form integral resulting from the superposition of focused waves.

All results here emphasized provide new signatures for the identification of disordered media; indeed the average DOP value and its histogram are microstructure-related and can be calibrated versus structural parameters of samples. Applications concern security and remote sensing, biophotonic and biomedical optics, lighting, microscopy and metrology.